\documentstyle[prb,aps,graphicx,multicol]{revtex}

\begin{document}
\draft

\title{Mesoscopic effects in tunneling between parallel quantum
wires}

\author{Daniel Boese$^1$, Michele Governale$^1$, Achim Rosch$^2$,
and Ulrich Z\"ulicke$^{1}$} 

\address{$^1$Institut f\"ur Theoretische Festk\"orperphysik,
Universit\"at Karlsruhe, D-76128 Karlsruhe, Germany\\
$^2$Institut f\"ur Theorie der Kondensierten Materie,
Universit\"at Karlsruhe, D-76128 Karlsruhe, Germany}

\date{\today}

\maketitle

\begin{abstract}

We consider a phase-coherent system of two parallel quantum wires
that are coupled via a tunneling barrier of finite length. The
usual perturbative treatment of tunneling fails in this case,
even in the diffusive limit, once the length $L$ of the coupling
region exceeds a characteristic length scale $L_t$ set by
tunneling. Exact solution of the scattering problem posed by the
extended tunneling barrier allows us to compute tunneling
conductances as a function of applied voltage and magnetic field.
We take into account charging effects in the quantum wires due to
applied voltages and find that these are important for
one-dimensional--to--one-dimensional tunneling transport.

\end{abstract}


% only for galley style output
\begin{multicols}{2}

% body of paper here
\narrowtext

\section{Introduction}

Tunneling provides a powerful tool to probe electronic properties
of matter.\cite{electun} Its sensitivity to momentum-resolved
spectral features is determined by geometrical details of the
tunnel junction. For example, this sensitivity is completely lost
when tunneling occurs via a point contact, whereas it is maximal
for an extended, clean tunneling barrier. Since the experimental
study of tunneling between two separately contacted, parallel,
vertically separated two-dimensional (2D) electron systems became
possible,\cite{weim:apl:88} electronic structure and interaction
effects in low dimensions have been the subject of careful
investigation. In the ideal case, conservation of canonical
momentum in the plane of the 2D electron systems leads to sharp
tunneling resonances; allowing for exploration of electronic
subband energies,\cite{smo:prl:89} mapping of the 2D Fermi
surface,\cite{jpe:prb:91} and life-time measurements of 2D
Fermi-liquid quasiparticles.\cite{sheena:prb:95} Modification of
one of the 2D layers into a superlattice of one-dimensional (1D)
quantum wires has been employed to measure vertical tunneling
between 1D and 2D electron systems.\cite{smo:prb:91} Constraints
on tunneling imposed by the requirement of simultaneous
conservation of energy and momentum can be tuned by the transport
voltage and external magnetic fields. In certain
situations,\cite{uz:prb-rc:96} this makes it possible to observe
features of the momentum-resolved single-electron spectral
function directly in tunneling transport. 

The method of cleaved-edge overgrowth\cite{cleaved1}
(CEO) makes it possible to create long and clean quantum wires
in GaAs/GaAlAs heterostructures.\cite{yac:prl:96}
Using the same technique, systems of two parallel quantum wires
with a high and extremely clean tunneling barrier between them
have been fabricated in double-layer structures.\cite{amir} This
opens up new possibilities for studying the peculiar dynamics of
electrons in interacting 1D systems\cite{voit:reprog:94,balents}
using 1D--to--1D tunneling.\cite{splitgate} In particular, both
the phase-coherence length and the elastic mean free path
$l_{\text{el}}$ for electrons in these quantum wires usually
exceed the wire length.\cite{yac:prl:00} This motivates the
present work where we analyze mesoscopic effects in 1D--to--1D
transport. In related contexts, phase-coherent transport in
double-wire systems was discussed in terms of device
applications. For example, a system of two parallel, identical
quantum wires coupled within a spatial region of length
$L<l_{\text{el}}$ via an adjustable tunneling
barrier\cite{dircoupl2} was proposed as a possible realization of
a current switch.\cite{dircoupl1} Wave packets of electrons
injected into one of the wires will be coherently transferred to
the other one and back with frequency $2 |t|/\hbar$. (Here we
denoted the tunnel-splitting  of energy levels in the coupling
region by $2 |t|$.) In steady state, this results in a coherent
charge oscillation in real space with wave length $L_t^{(0)}=\pi
\hbar v_{\text{F}}/|t|$. Modulation of $|t|$ controls the signal
at the output of the injecting wire. Ideally, it is maximal
(minimal) when the ratio of $L$ and $L_t^{(0)}$ is
(half-)integer. In reality, output characteristics depend
sensitively on details of the tunneling barrier.\cite{dircoupl3}
Assuming the feasibility to engineer barrier design, coupled
quantum wires were suggested\cite{qubit} as realizations of
quantum logical gates.
\begin{figure} 
\centerline{\includegraphics[width=7.5cm]{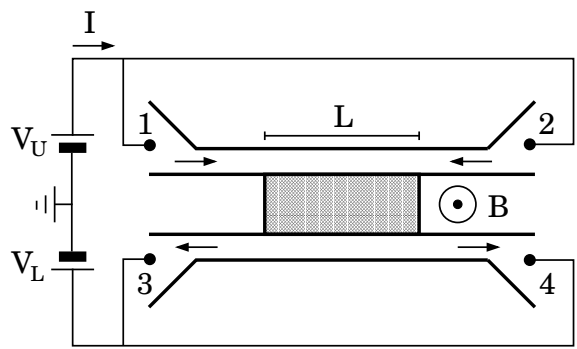}}
\vspace{0.3cm}
\caption{Schematic setup for a system of two parallel quantum
wires. The magnetic field $B$ allows tuning of kinetic vs.\
canonical momentum. A voltage $V_{\text{U(L)}}$ is applied
uniformly to the upper (lower) wire, i.e., raises the chemical
potential of {\em both\/} left-movers {\em and\/} right-movers.
The parts of the wires outside the region of space where the
barrier is finite are leads to ideal reservoirs. For simplicity,
we assume leads of infinite length in our model description.}
\label{fig:1}
\end{figure}

In this paper, we consider phase-coherent transport in a system
of parallel quantum wires coupled via a finite tunneling barrier.
See Fig.~\ref{fig:1}. Charging effects in the wires caused by
applied voltages influence tunneling in an important way
because they determine the degree to which 1D subbands are
shifted or filled. The basic physics of this interplay is
discussed in the
following section. Our microscopic model for the double-wire
system is introduced in the first part of Sec.~\ref{model}. Apart
from capacitance effects, interactions are neglected within our
approach, which is therefore valid only for voltages and in-plane
magnetic fields probing the 1D electron systems beyond the
cut-off for Luttinger-liquid behavior.\cite{voit:reprog:94}
Results from lowest-order perturbation theory are compared with
the exact solution using scattering theory. We calculate linear
and differential conductances for 1D--to--1D tunneling transport,
and discuss their features in Sec.~\ref{discuss}.
  
\section{Effect of an applied voltage}

In the typical tunneling experiment, a voltage drop $V$ across
the barrier drives a current. Microscopically, it is often
assumed that the voltage shifts quasiparticle bands in the two
subsystems by $\pm e V/2$, respectively, as compared to the
equilibrium situation where no net current flows. The external
circuit is supposed to prevent charging of the
subsystems.\cite{sca:ann:74} In general, however, the applied
voltage will shift the bands as well as partly fill them. As the
I--V curve for 1D--to--1D tunneling depends sensitively on the
scenario of band filling vs.\ band shifting, we discuss this
issue here in some detail.

At zero temperature, the free energy (per length) of a 1D system
is given by its total energy (per length) $E_{\text{tot}}$ which
is a functional of particle density $n$. In a clean quantum wire,
$n=n_0$ will be constant. Before applying a voltage, the system
is assumed to be charge-neutral, i.e., the uniform electronic
charge density $e n_0$ is compensated by positive background and
image charges. It is useful to divide $E_{\text{tot}}$ into two
parts; $E_{\text{tot}}=E_{\text{int}}+E_{\text{Coul}}$. All
Coulombic terms (including the Hartree energy of electrons in the
wire) are collected in $E_{\text{Coul}}$, and $E_{\text{int}}$ is
the internal energy of the quantum wire comprising kinetic and
exchange-correlation contributions. For our purposes, we adopt
the simple model with $E_{\text{Coul}}=(e\Delta n)^2/2 \tilde C$
where $\Delta n$ is the deviation from the neutralized charge
density $n_0$, and $\tilde C$ denotes the electrostatic
capacitance per unit length of the wire.\cite{conleads} The
applied voltage is assumed to lead to a uniform shift $e V$ with
respect to the equilibrium chemical potential $\mu_0=\left.
\partial E_{\text{tot}}/\partial n \right|_{n_0}$ of the
wire.\cite{voltcaveat} The induced change $\Delta n$ in the total
density has to be calculated from
\begin{equation}\label{genvolt}
\mu_0 + e V=\frac{e^2}{\tilde C}\,\Delta n+\left.\frac{\partial
E_{\text{int}}}{\partial n}\right|_{n=n_0+\Delta n} \quad .
\end{equation}
In the limit of small voltages ($|e V| \ll \mu_0$) where
linear-response theory is valid, we can use
\begin{figure} 
\centerline{\includegraphics[width=7.5cm]{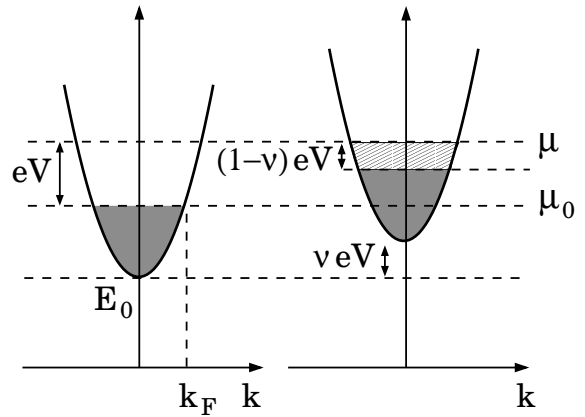}}
\vspace{0.3cm}
\caption{A voltage $V$ applied to a quantum wire results, in
general, both in a uniform shift of the 1D subband and charging
of the wire. Outside the linear-response regime, the parameter
$\nu$ depends on voltage. In the double-wire system, a
self-consistent treatment of charging effects due to the voltage
and tunneling is necessary unless tunneling is weak.}
\label{fig:2}
\end{figure}
\begin{equation}
\left.\frac{\partial E_{\text{int}}}{\partial n}\right|_{n=n_0+
\Delta n} - \mu_0 \approx \frac{\Delta n}{D_0} \quad .
\end{equation}
Here, $D_0$ is the thermodynamic\cite{thdydos} density of states 
(DOS) defined by $D_0=\left.\partial n/\partial\mu
\right|_{\mu=\mu_0}=\big(\partial^2 E_{\text{int}}/\partial n^2
\big)^{-1}_{n=n_0}$. In the linear-response limit, it is then
possible to express $\Delta n$ explicitly in terms of the
external voltage,\cite{buttbefore}
\begin{equation}
\label{eq:dens}
\Delta n = e V \,\, \frac{D_0}{1+\zeta}\quad ,
\end{equation}
where the parameter $\zeta$ measures the relative importance of
Coulombic and density-of-states effects:
\begin{equation}
\label{eq:zeta}
\zeta = \frac{e^2}{\tilde{C}}\, D_0 \quad .
\end{equation}
We see that, for $\zeta \ll 1$, a voltage will simply {\em fill\/}
quasiparticle bands without shifting them (band-filling limit).
In particular, this applies in the absence of electron-electron
interactions. In the opposite case $\zeta \gg 1$, an applied
voltage shifts the bands (band-shifting limit). This situation is
analogous to that of a bulk metal or a single-electron
transistor.\cite{set} To
get an idea of the situation realized in our system of interest,
we estimate the capacitance of a quantum wire by $\tilde{C} = 2
\pi\epsilon / \ln(R/r)$, with $R$ being the distance to
surrounding metal gates, and $r$ denoting the characteristic
transverse dimension of the wire. Typical values are $\sim
10^{-10}$~F/m. With Fermi energies of quantum wires ranging
between $1\dots 10$~meV, we obtain $\zeta\sim 1\dots 10$. Hence,
typical quantum wires are in the intermediate regime where both
band-filling and band-shifting occurs at the same time. This case
is illustrated in Fig.~\ref{fig:2}. It is important to keep in
mind, however, that Eqs.~(\ref{eq:dens}) and (\ref{eq:zeta}) are
only valid when $\Delta n\ll n_0$. In experiment, voltages
comparable to and larger than $\mu_0$ are applied to probe the
full single-particle dispersion relation.\cite{amir} Then, for a
quantitative comparison between theory and experiment, $\Delta n$
has to be found from Eq.~(\ref{genvolt}). For example, a wire
whose density was initially large enough for it to be in the
band-filling limit crosses over to the band-shifting limit when
it is depleted.

At this point, it is useful to make contact with results obtained
for Tomonaga-Luttinger (TL) models\cite{tom:prog:50}
of interacting 1D electron systems. Unlike their
higher-dimensional counterparts, 1D metals cannot be described
within the Fermi-liquid paradigm. Instead, their low-energy
properties are represented by effective TL models, and the
phenomenology of a Luttinger liquid\cite{fdmh:jpc:81} (LL)
applies. Instead of Landau parameters, it is the velocities of
certain collective and zero-mode excitations that determine all
physical quantities of a LL. In particular, the ratio
$r_{\text{N}}=v_{\text{N}}/v_{\text{F}}$ of the velocity
$v_{\text{N}}$ of the charged zero mode\cite{fdmh:jpc:81} and the
bare Fermi velocity enters the expression for the electrostatic
capacitance per unit length of a Luttinger liquid: $\tilde
C_{\text{LL}}=e^2 D_0/(r_{\text{N}}-1)$. Here, $D_0=1/\pi\hbar
v_{\text{F}}$ is the 1D DOS. Using Eq.~(\ref{eq:zeta}), we find
$\zeta_{\text{LL}}=r_{\text{N}}-1$. The non-interacting case
where $r_{\text{N}}=1$ corresponds to the band-filling  limit,
whereas strong Coulomb interactions ($r_{\text{N}}\to\infty$)
recover the band-shifting limit. We would like to remark that
$r_{\text{N}}$ constitutes an independent parameter in the
low-energy theory of any given real quasi-1D system. It is
unrelated, except in certain special cases,\cite{slava:prl:98} to
the famous interaction parameter $K_\rho$ that enters power-law
expressions for electronic correlation
functions.\cite{voit:reprog:94} From now on, we consider the
model where $r_{\text{N}}>1$ but $K_\rho=1$. This approximation
is valid to describe current experiments where the wires are
probably not long enough for the power-law characteristics of a
LL to be observable.\cite{LLcaveat1} Even for infinitely long
wires, however, our results apply at energies and wave vectors
far enough from the Fermi points where the single-particle
spectral function recovers Fermi-liquid-like
characteristics.\cite{med:prb:93b}

\section{Model and Formalism}
\label{model}

We consider two quantum wires of infinite length, labeled U(pper)
and L(ower), that are parallel to the $x$ direction and located,
in the $yz$ plane, at $y=0$ and $z=z_{\text{U,L}}$. The potential
barrier between them is assumed to be finite and uniform in the
region $|x|\le L/2$ and infinite otherwise. Within the standard
notation of second quantization, the Hamiltonian for our system
is given by
\begin{mathletters}\label{hamiltonian}
\begin{eqnarray}
H &=& \sum_{\alpha=\text{U,L}} H_{\alpha}+ H_{\text{tun}}\quad ,
\\ H_{\alpha}&=&\int\!\frac{d k}{2\pi}\,\, \epsilon_\alpha (k)\,
c_{k\alpha}^\dagger c_{k\alpha}^{}\quad ,\\
H_{\text{tun}} &=& \int_{-L/2}^{L/2}\!\! dx \left\{ t\,
\psi^\dagger_{\text{U}}(x)\psi_{\text{L}}(x) + {\mathrm{H.c.}}
\right\} \\
&=&\int\!\frac{d k}{2\pi}\int\!\frac{d p}{2\pi}\,\,\left\{t_{k,p}
\, c_{k\text{U}}^\dagger c_{p\text{L}}^{} + \,\text{H.c.}\right\}
\quad .
\end{eqnarray}
\end{mathletters}
Here, $\epsilon_{\text{U(L)}}(k)$ is the electronic dispersion
relation in the upper (lower) wire. Modulo an unimportant phase
factor, the tunneling matrix element is given
by\cite{gov:prb:00b}
\begin{equation}\label{tkp} 
t_{k,p}=2 |t|\,\frac{\sin\left[(p-k)L/2\right]}{p-k} = |t|\,
\sqrt{2\pi L\,\delta_{L}(p-k)}.
\end{equation}
The second equality in Eq.~(\ref{tkp}) constitutes the definition
of $\delta_{L}(p-k)$ which is a finite-size realization of
Dirac's $\delta$-function. Tunneling  occurs mainly between
states with momenta satisfying $|p-k|<2\pi/L$. Perfect momentum
conservation holds only in the limit $L\rightarrow\infty$.

We consider the case where a single (the lowest) 1D subband in
each wire is occupied and assume a parabolic subband dispersion.
The effect of a magnetic field $\vec B = B\,\hat y$ applied
perpendicularly to the plane of the two wires can be included by
a shift of kinetic with respect to canonical
momentum.\cite{magcaveat} Then, dispersion relations read
($\alpha=\text{U,L}$)
\begin{equation}\label{eq:dispersions}
\epsilon_\alpha(k)=\frac{\hbar^2}{2m}\left(k-\frac{e B}{\hbar}
\,z_\alpha\right)^2+E_{0\alpha}+\nu_\alpha\, e V_\alpha \quad .
\end{equation}
Here, $m$ is the effective electron mass in the semiconductor
host medium, and $E_{0\alpha}$ denotes the energy at the bottom
of the respective wire's lowest 1D subband. The term
$\nu_{\text{U(L)}}e V_{\text{U(L)}}$ takes into account the
shifting of the band in the upper (lower) wire due to an applied
voltage. See Fig.~\ref{fig:2}. For simplicity, we neglect
effects due to the mutual capacitance of the two wires, which can
be included straightforwardly. In general, the values of
$\nu_\alpha$ will depend on voltage. Furthermore, except in the
limit of weak tunneling where $|t|\ll |e V_\alpha|$, they have to
be determined from a self-consistent treatment of charging
effects resulting from tunneling and electrostatics. While this
is, in principle, straightforward to implement, we choose to
focus here on the weak-tunneling limit which is more relevant for
current experiment.\cite{amir} In the linear-response regime, we
have $\nu_\alpha=\zeta_\alpha/(1+\zeta_\alpha)$ with
$\zeta_\alpha$ defined for each wire in analogy to
Eq.~(\ref{eq:zeta}).

Absolute values of energy and the $z$ coordinate are irrelevant;
results depend only on the difference of subband energies,
$\Delta E_0=E_{\text{0L}}-E_{\text{0U}}$, and the wire
separation, $d=z_{\text{U}}-z_{\text{L}}$. For simplicity, we
choose $E_{\text{0U}}=0$ and $z_{\text{U}}=0$ in the following.
Also, to avoid cluttering the notation, we have suppressed spin
quantum numbers. In typical CEO structures, the effect of Zeeman
splitting is negligible for the range of magnetic fields to be
considered below.\cite{zeecaveat} Hence, electron spin leads only
to factors of 2 which we include in our final formulae for
tunneling current and conductances.

\subsection{Perturbation theory: Lowest order in tunneling}

A standard procedure\cite{sca:ann:74,mahan} for calculating the
tunneling current is to perform \pagebreak perturbation theory in
$H_{\text{tun}}$. To leading order, the current flowing
from the upper to the lower wire is
\begin{eqnarray}
I&=& \frac{2 e}{\hbar} \int\! \frac{dk}{2\pi}\int\! \frac{d p}{2
\pi}\,\, |t_{k,p}|^2 \int_{-\infty}^{\infty} \!\frac{d\epsilon}{2
\pi}\,\, A_{\text{U}}(\epsilon,k)A_{\text{L}}(\epsilon,p)
\nonumber \\ &&\hspace{4cm}\times
\left[ f_{\text{U}}(\epsilon)-
f_{\text{L}}(\epsilon )\right] \quad , \label{perturb}
\end{eqnarray}
with Fermi functions $f_\alpha(\epsilon)=1/[1+\exp\{(\epsilon-
\mu_0-e V_\alpha)/k_{\text{B}}T\}]$. The single-particle spectral
functions for the wires are given, within the model specified
above, by $A_\alpha(\epsilon,k) = 2 \pi \delta[\epsilon -
\epsilon_\alpha(k)]$. In the linear-response limit ($|e V_\alpha|
\ll\mu_0$), we find the tunneling conductance
\begin{equation}\label{eq:pert1}
G =\frac{2e^2}{\hbar^3}\frac{|t|^2 L}{v_{\text{FU}}v_{\text{FL}}}
\sum_{\gamma,\gamma^\prime=\pm 1}\delta_{L}\left(\frac{\pi}{2}[
\gamma n_{\text{U}}-\gamma^\prime n_{\text{L}}]-p_{\text{B}}
\right)\, .
\end{equation}
Here, $v_{\text{F}\alpha}$ and $n_{\text{F}\alpha}$ denote the 
Fermi velocity and electron density of the respective wire at
the equilibrium chemical potential $\mu_0$. The peak-shape
function $\delta_L$ has been defined in Eq.~(\ref{tkp}), and the
relative shift of the 1D Fermi seas due to the applied magnetic
field is $p_{\text{B}} = - e B\, d/\hbar$. 

In analogy to 2D--to--2D tunneling,\cite{jpe:prb:91} resonances
appear in the tunneling conductance $G$ as function of magnetic
field whenever parts of the shifted Fermi surfaces of the two
wires overlap. As the 1D Fermi surface consists of just two
points, the shape of these resonances is that of a smeared delta
function of width $2\pi/L$. The peak value of the tunneling
conductance can be written as
\begin{equation}\label{eq:peak}
G_{\text{pk}} = n^*\,\frac{2 e^2}{h}\,\,\left(\frac{\pi L}{L_t}
\right)^2 \quad ,
\end{equation} 
where $L_t = \pi\hbar\sqrt{v_{\text{FU}}v_{\text{FL}}}/|t|$ is an
effective length scale introduced by tunneling, and $n^*=1$ or
$2$ depending on the number of overlapping Fermi points at peak
condition.

In the dirty limit\cite{dirtnote} where the length $L$ of the
tunneling barrier is larger than the mean free path
$l_{\text{el}}$, Eq.~(\ref{eq:pert1}) is still valid but the
peak width is now given by $2\pi/l_{\text{el}}$. Also, the factor
$(\pi L/L_t)^2$ in Eq.~(\ref{eq:peak}) has to be replaced by $2
\pi^2 L\,l_{\text{el}}/L^2_t$. In both the ballistic and
diffusive cases, taking the limit of $L\to\infty$ is unphysical:
the conductance through the barrier cannot exceed $2 e^2/h$ per
channel. Hence, the actual small parameter enabling perturbative
treatment of tunneling is $L/L_t$. Smallness of $L/L_t$ means
that the time between tunneling events has to be larger than the
time it takes electrons to traverse the region where the
potential barrier between the wires is finite. Only then it will
be possible to neglect higher-order effects due to electrons
tunneling coherently back and forth between the wires. Using the
exact solution developed in the next subsection, we will find
indeed that the perturbative result displayed in
Eq.~(\ref{eq:pert1}) is valid only as long as $L\ll L_t$.

\subsection{Exact solution using scattering theory}

\begin{figure} 
\centerline{\includegraphics[width=7.5cm]{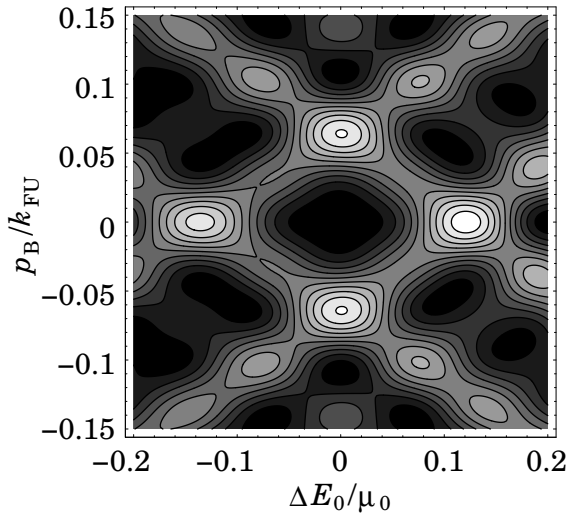}}
\vspace{0.3cm}
\caption{Contour plot of the linear conductance $G$ vs.\ $\Delta
E_{\mathrm{0}}$  and $p_{\mathrm{B}}$ for $L=L_t^{(0)}=100/
k_{\text{F}}^{(0)}$. (Here, $L_t^{(0)}$ and $k_{\text{F}}^{(0)}$
are the tunneling length and Fermi wave vector when $\Delta E_0=
0$ and $B=0$. Due to our gauge choice, $k_{\text{FU}}=\pi
n_{\text{U}}/2$ is unaffected by the external magnetic field.)
The oscillatory structure is well-described by
Eq.~(\ref{approxcond}). For our choice of parameters, $G=0$ at
in-resonance condition. Note that tuning the magnetic field near
off-resonance maxima (e.g., for $\Delta E_0=0$ and $p_{\text{B}}=
0.07\, k_{\text{FU}}$) has a strong effect on $G$.}
\label{fig:4}
\end{figure}

As the model defined in Eqs.~(\ref{hamiltonian}) describes two
systems of noninteracting fermionic quasiparticles that are 
coupled via tunneling in a finite region of space, we can use
scattering theory for calculating
transport.\cite{landauer2} To make
this explicit, we rewrite the Hamiltonian of our system in
first-quantized notation and real-space representation. It is a
$2\times 2$ matrix [because wave functions are two-component
spinors $(\psi_{\text{U}},\psi_{\text{L}})^{\mathrm{T}}$]:
\begin{equation}\label{Hmat}
H = \left( \begin{array}{lr} \epsilon_{\text{U}}(-i\partial_x) &
t(x) \\ t(x) & \epsilon_{\text{L}}(-i\partial_x) \end{array}
\right) \quad .
\end{equation}
The tunneling matrix element is piecewise constant: $t(x)=|t|$
for $|x|\le L/2$ and $t(x)=0$ otherwise. Hence, regions with
$|x|>L/2$ where the wires are independent act as leads where
scattering states can be defined. We attach labels 1 through 4 to
these leads as shown in Fig.~\ref{fig:1}. The region $|x|\le
L/2$ where tunneling occurs acts as an effective scatterer.
The current flowing through the tunnel barrier is then given by
\begin{equation}\label{curr1}
I=\frac{2 e}{h}\sum_{m=1,2 \atop n=3,4} \int d \epsilon\,\left|
T_{m,n}(\epsilon)\right|^2\left[ f_{\text{U}}(\epsilon) -
f_{\text{L}} (\epsilon)\right]\quad ,
\end{equation}
where $T_{m,n}(\epsilon)$ denotes the transmission coefficient
for electrons with energy $\epsilon$ that originate in lead $m$
and are scattered into lead $n$. We calculate the transmission
coefficients by matching scattering states in the leads to the
appropriate eigenstates of the Hamiltonian~(\ref{Hmat}) in the
region $|x|\le L/2$. As this is a straightforward exercise, and
results for the most general case are lengthy, we omit explicit
formulae here. Due to the difference of Fermi functions in
Eq.~(\ref{curr1}), only transmission coefficients $T_{m,n}
(\epsilon)$ at energies within the voltage window, i.e., with
$\epsilon-\mu_0\in [e V_{\text{L}}, e V_{\text{U}}]$,
contribute. In the limit of small applied voltage,
Eq.~(\ref{curr1}) yields the linear conductance 
\begin{figure} 
\centerline{\includegraphics[width=7.5cm]{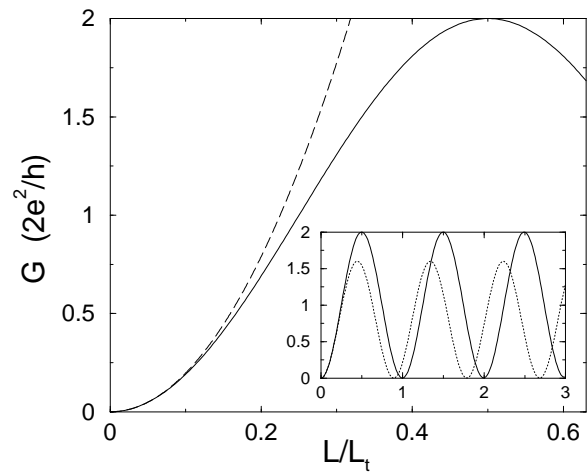}}
\caption{Linear conductance for two identical wires, calculated
exactly (solid curve) and with perturbation theory (dashed
curve). Inset: Oscillation of conductance in zero magnetic field
(solid) and with $p_{\text{B}}=0.001\, k_{\text{F}}^{(0)}$
(dotted) where $k_{\text{F}}^{(0)}$ is the Fermi wave vector for
zero magnetic field.}
\label{fig:3}
\end{figure}
\begin{equation}\label{linex}
G=\frac{2 e^2}{h}\sum_{m=1,2\atop n=3,4}\left|T_{m,n}(\mu_0)
\right|^2 \quad .
\end{equation}
As expected, $G$ obtained from Eq.~(\ref{linex}) deviates from
the perturbative result [Eq.~(\ref{eq:pert1})] for long enough
barrier length $L$, see Fig~\ref{fig:3}. The oscillatory
dependence of $G$ on $L$ can be tuned by the applied magnetic
field, as seen in the inset of Fig.~\ref{fig:3}. When the
effective tunnel splitting in the coupling region is much smaller
than the Fermi energy of the quantum wires, the following
approximate formula for the linear conductance can be derived
(see the Appendix for details):
\begin{equation}\label{approxcond}
G\approx\frac{2 e^2}{h}\sum_{\gamma,\gamma^\prime=\pm 1}
\frac{\sin^2\left(\pi\sqrt{\left(L/L_t\right)^2 + \left(L/
L_{\gamma\gamma^\prime}\right)^2}\right)}{1+(L_t/L_{\gamma
\gamma^\prime})^2}\quad .
\end{equation}
Here, new length scales $L_{\gamma\gamma^\prime}$ appear that
measure the mismatch of canonical Fermi momentum for pairs of
Fermi points from the upper (right-mover $\gamma=+1$, left-mover
$\gamma=-1$) and lower ($\gamma^\prime$ analogous) wires:
\begin{equation}\label{reslength}
L_{\gamma\gamma^\prime}=\frac{2\pi}{\frac{\pi}{2}\left[\gamma
n_{\text{U}}-\gamma^\prime n_{\text{L}}\right] - p_{\text{B}}}
\quad .
\end{equation}
Exact calculation of $G$ in the appropriate limit confirms the
validity of Eq.~(\ref{approxcond}); see Fig.~\ref{fig:4}.

\section{Results and Discussion}
\label{discuss}

Different regimes in the behavior of the linear 1D--to--1D
tunneling conductance are distinguished by the interplay of the
relevant length scales encountered above. These are the length
$L$ of the tunnel barrier, $L_t$ which is a measure of the
strength of tunneling, and lengths $L_{\gamma\gamma^\prime}$
which are defined for any pairing of a Fermi point from the upper
wire with one from the lower wire. For the following discussion,
we consider only the largest $L_{\gamma\gamma^\prime}$ of all
possible. Comparing $L_t$ with the other lengths, a
weak-tunneling regime ($L_t>\max\{L_{\gamma\gamma^\prime},L\}$)
can be distinguished from a strong-tunneling regime ($L_t<\max\{
L_{\gamma\gamma^\prime},L\}$). Furthermore, we call the system
{\em in resonance\/} when the Fermi points of a corresponding
$L_{\gamma\gamma^\prime}$ are close to each other on the scale of
$2\pi/L$, i.e., when $L<L_{\gamma\gamma^\prime}$.
Conversely, the off-resonance limit is reached for $L>
L_{\gamma\gamma^\prime}$.

In the strong-tunneling regime, the linear conductance oscillates
as a function of $L$ with wave length $L_t$ and maximum amplitude
$2 e^2/h$ ($4 e^2/h$ for identical wires). Previously, when the
feasibility of using the double-wire system as a directional
coupler was discussed, the in-resonance limit was considered
only.\cite{dircoupl1} Control of directional-coupler operation is
then possible only by varying $L_t$, i.e., essentially only by
adjusting the barrier height. Here we find that, in the
off-resonance limit, the device is tunable, in addition to
varying $t$, by an applied magnetic field or, equivalently, by
adjusting the density mismatch in the two wires. This is seen
already in the inset of Fig.~\ref{fig:3} and, more clearly, in
Fig.~\ref{fig:4}. It is, therefore, possible to adjust the
effective length scale for coherent electron transfer between the
wires by applying a magnetic field. When $L$ is equal to
several times $L_t$, the difference between the effective
transfer length in a magnetic field and $L_t$ leads to an
accumulated phase shift over many oscillations that can reach
$\pi/2$ without concomitant loss in amplitude. In particular for
CEO structures, operating the system in off-resonance mode
provides a convenient alternative to any (hardly feasible)
adjustment of the high tunnel barrier.

The weak-tunneling limit is well-suited for spectroscopic
application of 1D--to--1D tunneling. Sharp peaks are exhibited by
both the linear and differential conductances for in-resonance
conditions. Measured on the scale of resonance peaks, the
off-resonance conductance is orders of magnitude smaller. We have
calculated the differential conductance as a function of magnetic
field and voltage whose resonance condition corresponds to a
Fermi point of one wire coinciding with a point on the dispersion
curve of the other wire. The exact location of these coincidences
in the $V$--$B$ plane depends sensitively on charging effects in
the quantum wires. In the following, we focus on the limit of
weakly coupled wires where the self-consistent charge profile is
not importantly affected by tunneling. Furthermore, we consider
the situation with symmetric bias $V_{\text{U}}=- V_{\text{L}}=V/
2$.
\begin{figure} 
\centerline{\includegraphics[width=7.5cm]{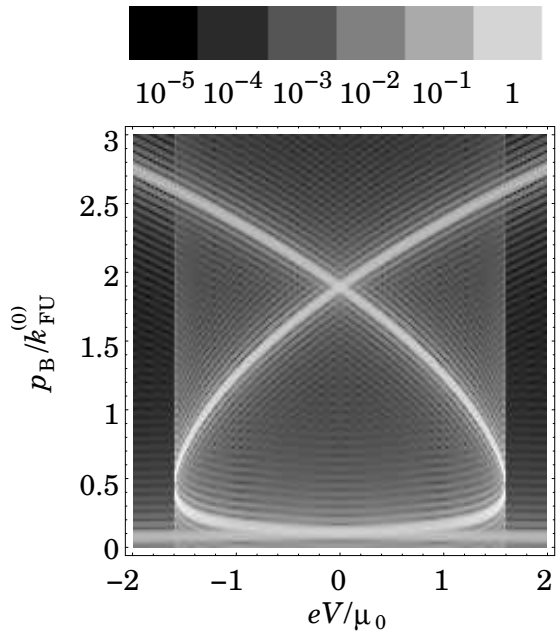}}
\vspace{0.3cm}
\caption{Differential conductance for 1D--to--1D tunneling in the
ideal band-filling case ($\tilde C=\infty$), shown as a
logarithmic gray-scale plot in arbitrary units. Parameters used
are $\Delta E_0=0.2\mu_0$ and $k_{\text{FU}}^{(0)} L = 100$.
$k_{\text{FU}}^{(0)}$ is the Fermi wave vector of the upper wire
at zero voltage. The finite width of bright resonance features as
well as the appearance of darker maxima is due to the finite
length $L$ of the tunnel barrier. Note the symmetry with respect
to voltage reversal which is a key feature of the band-filling
case.}
\label{fig:5}
\end{figure}
\begin{figure} 
\centerline{\includegraphics[width=7.5cm]{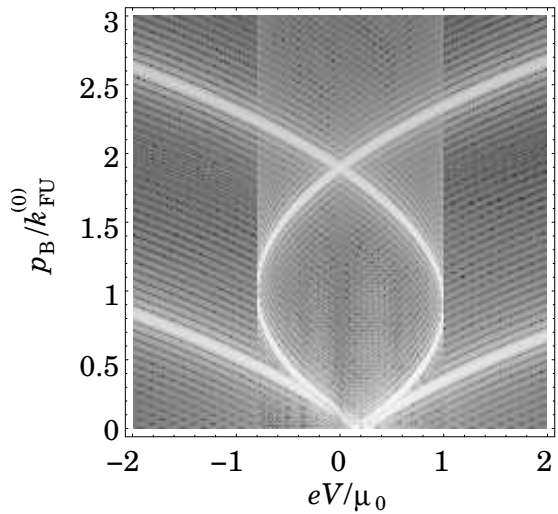}}
\vspace{0.3cm}
\caption{Differential conductance for 1D--to--1D tunneling in the
ideal band-shifting case ($\tilde C=0$). We show a logarithmic
gray-scale plot of its absolute value. Parameters and gray-scale
units are the same as in Fig.~\ref{fig:5}. A characteristic
feature of the band-shifting case is that the edges of the
leaf-shaped structure in the voltage direction are exactly at the
Fermi energies of the two wires. The differential conductance is
negative on the low-magnetic-field resonance lines.}
\label{fig:6}
\end{figure}
\begin{figure}
\centerline{\includegraphics[width=7.5cm]{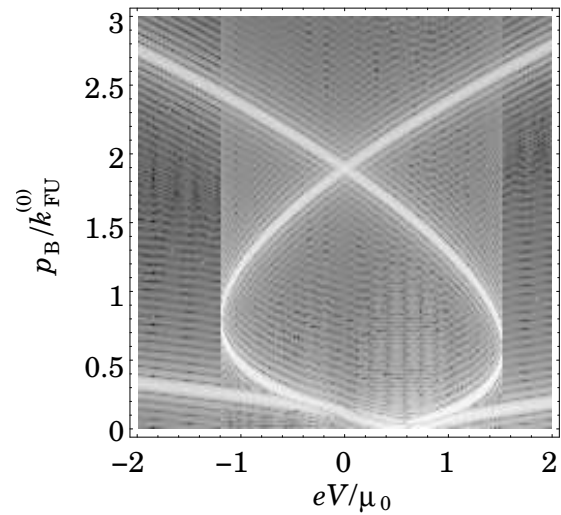}}
\vspace{0.3cm}
\caption{Differential conductance for 1D--to--1D tunneling in an
intermediate situation with finite $\tilde C$. Shown is its
absolute value in a logarithmic gray-scale plot. See
Fig.~\ref{fig:5} for a legend. In addition to the parameters used
in Fig.~\ref{fig:5}, we have $\tilde C_{\text{U}}=
\tilde C_{\text{L}} = 8 \varepsilon\varepsilon_0/
(k_{\text{FU}}^{(0)} a_{\text{B}})$ where $a_{\text{B}}$ is the
Bohr radius in the semiconductor host material.}
\label{fig:7}
\end{figure}
Figures~\ref{fig:5} and \ref{fig:6}, respectively, show 
logarithmic gray-scale plots of the absolute value of the
differential tunneling conductance for the ideal band-filling
($\tilde C=\infty$) and band-shifting ($\tilde C=0$)
cases.\cite{pbunits} Bright lines are formed by points in the
$V$--$B$ plane where the above-mentioned resonance condition is
fulfilled. Due to the finite length of the tunnel barrier, more
maxima appear with peak values being orders of magnitude smaller
than at the resonance peaks. In the band-shifting case, resonance
lines are direct images of parts of the wires' electronic
dispersion curves. In particular, the extension of the
leaf-shaped structure in the positive and negative voltage
direction provides a direct measure of the respective wire's
Fermi energy. This is not the case for the band-filling limit,
which is characterized by a resonance line running close to the
voltage axis when $E_0\ll \mu_0$. Its leaf structure is
symmetric under voltage reversal, with extension in (positive or
negative) voltage direction given by $2 (\mu_0 - |\Delta E_0|)$.
In real systems, the capacitance $\tilde C$ is finite, and an
intermediate picture will be obtained for the differential
tunneling conductance. An example is shown in Fig.~\ref{fig:7},
where $\tilde C$ has a value such that $\zeta_{\text{U}}=0.5$ at
zero voltage. Depletion of one of the wires for increasing
voltage leads to a cross-over to the band-shifting situation. As
a result, the ideal band-shifting limit is not easily
distinguished from the intermediate case. Quantitative comparison
of the measured resonance structure in the differential
conductance with results expected from an independent measurement
of Fermi energies, electron densities etc.\ will have to include
the effect of the finite $\tilde C$. Conversely, for known
Fermi-sea parameters, the value of $\tilde C$ can be extracted by
fitting the measured resonance pattern of the differential
conductance for 1D--to--1D tunneling.

\section{Conclusions}
Motivated by recent experiment, we have investigated linear and
differential conductances for 1D--to--1D transport. Our results
show that effects due to phase coherence and charging of the
wires are important for realistic double-wire structures. Regimes
of weak and strong tunneling, as well as in and out of resonance,
are distinguished and their key features discussed. We point out
new possibilities for device application of 1D--to--1D tunneling
and its use for electron-dispersion spectroscopy.

\acknowledgments
Useful discussions with M.~B\"uttiker, G.~Sch\"on, and A.~Yacoby
are gratefully acknowledged. This work was supported by DFG
within Sonderforschungsbe\-reich 195, Graduiertenkolleg
"Kollektive Ph\"anomene im Festk\"orper" (D.B.), and the
Emmy-Noether-Programm (A.R.). U.Z.\ thanks the Braun Submicron
Center at the Weizmann Institute, Israel, for hospitality during
a visit sponsored by the EU LSF programme. 

\appendix
\section*{Calculation of transmission coefficients}

Equations~(\ref{curr1}) and (\ref{linex}) express tunneling
current and linear conductance in terms of transmission
coefficients $T_{m,n}(\epsilon)$. These transmission coefficients
can be obtained exactly by matching eigenstates of Hamiltonian
$H$ [given in Eq.~(\ref{Hmat})] with eigenvalue $\epsilon$ in
the coupling region ($|x|\le L/2$) to appropriate eigenstates in
the leads. For example, to calculate $T_{1,n}$, we use the ansatz
\begin{mathletters}\label{wfansatz}
\begin{eqnarray}
\left.\Psi_\epsilon(x)\right|_{x<-\frac{L}{2}} &=& \left(
\begin{array}{c} 1\\0\end{array}\right) e^{i k_{\text{U}}^{(+)} x
}\nonumber \\ &+& \left(\begin{array}{c} t_{11}\\0
\end{array}\right) e^{i k_{\text{U}}^{(-)} x}+\left(\begin{array}
{c}0\\ t_{13} \end{array}\right)e^{i k_{\text{L}}^{(-)} x} \, , 
\\ \left.\Psi_\epsilon(x)\right|_{x>\frac{L}{2}} &=& \left(
\begin{array}{c} t_{12} \\0\end{array}\right) e^{i
k_{\text{U}}^{(+)} x} +\left(\begin{array}{c}0\\t_{14}\end{array}
\right) e^{i k_{\text{L}}^{(+)} x}\, , \\
\left.\Psi_\epsilon(x)\right|_{|x|<\frac{L}{2}} &=& \sum_{\alpha=
\text{a,b}} \left\{d_+^{(\alpha)}\left(\begin{array}{c}
u_+^{(\alpha)}\\ v_+^{(\alpha)}\end{array}\right)e^{i k_{+}^{
(\alpha)} x}\right. \nonumber \\ && \hspace{1cm} + \left.
d_-^{(\alpha)}\left(\begin{array}{c}u_-^{(\alpha)}\\
v_-^{(\alpha)}\end{array}\right)e^{i k_{-}^{(\alpha)} x}\right\}
\, .
\end{eqnarray}
\end{mathletters}
Here, wave vectors $k_{\text{U(L)}}^{(\pm)}$ are solutions of
$\epsilon\equiv\epsilon_{\text{U(L)}}(k_{\text{U(L)}}^{(\pm)})$
with positive ($+$) and negative ($-$) group velocity
$v_{\text{U(L)}}=\partial_{\hbar k}\epsilon_{\text{U(L)}}(k)$,
respectively. With the energy dispersion in the coupling region
given by $\epsilon_{\pm}(k)=\frac{1}{2}\left[\epsilon_{\text{U}}
(k)+\epsilon_{\text{L}}(k)\right]\pm |t| \sqrt{1+r^2}$, we have
$\epsilon\equiv\epsilon_{\pm}\big(k_{\pm}^{(\text{a,b})}\big)$.
The function $r=[\epsilon_{\text{U}}(k)-\epsilon_{\text{L}}(k)]/2
|t|$ measures the mismatch in the dispersions of the two wires
and determines the amplitudes $u_\pm=\sqrt{(1\pm r/\sqrt{1+r^2})/
2}$, $v_\pm=\pm\sqrt{(1\mp r/\sqrt{1+r^2})/2}$. Requiring
continuity of the wave function and current conservation at the
locations $x=\pm L/2$ yields a system of linear equations from
which the coefficients $t_{mn}$ and $d_\pm^{(\alpha)}$ are found.
Transmission coefficients entering Eqs.~(\ref{curr1}) and
(\ref{linex}) are then given by $T_{m,n}(\epsilon)=t_{mn}\sqrt{
|v_{\text{L}}/v_{\text{U}}|}$.

The linear conductance is given in terms of transmission
coefficients\cite{landauer2} at the equilibrium chemical
potential $\mu_0$, as expressed in Eq.~(\ref{linex}). To derive
Eq.~(\ref{approxcond}), we consider, e.g., $T_{1,4}$. When the
tunnel splitting $2|t|$ and Fermi-energy mismatch of the two
wires is much smaller than their Fermi energies, only amplitudes
for right-moving partial waves in ansatz~(\ref{wfansatz}) are
significantly different from zero. Neglecting left-moving partial
waves and the small density mismatch, the matching procedure
yields $t_{14}=i\sin(k_t L)/\sqrt{1+r^2}$ with $k_t=t\sqrt{1+r^2}
/\hbar v_{\text{F}}$. In the regime considered, we have $r\approx
L_t/L_{11}$ with $L_{11}$ defined in Eq.~(\ref{reslength}).
Similar calculations for other transmission coefficients finally
yield Eq.~(\ref{approxcond}).

\end{multicols}
 

\begin{thebibliography}{10}
\bibitem{zeecaveat}
The range of magnetic fields that need to be applied to map out
the interesting features in the linear, or differential,
tunneling conductance implies the condition $g\, m/m_0\ll n \, d$
for safe neglect of Zeeman splitting. Here, $g$ is the Land\'e
factor, $m_0$ the electron mass in vacuum, and $n$ the (in
general, voltage-dependent) electron density in the wire.
Typically, this condition will be violated only in the
band-filling limit around the point where one of the wires is
empty.
\bibitem{dirtnote}
Elastic scattering by impurities is taken into account by
introducing a finite life-time broadening for the single-electron
spectral functions $A_\alpha(\epsilon,k)$.
\bibitem{pbunits}
In Figs.~\ref{fig:5}--\ref{fig:7}, we give $p_{\text{B}}$ in
units of $k_{\text{FU}}^{(0)}\equiv \pi\, n_{\text{U}}^{(0)}/2$
which is independent of magnetic field due to our gauge choice.
\end{thebibliography}
\end{document}